\documentclass[10pt,conference]{IEEEtran}
\IEEEoverridecommandlockouts
\usepackage{cite}
\usepackage{amsmath,amssymb,amsfonts}
\usepackage{algorithmic}
\usepackage{graphicx}
\usepackage{textcomp}
\usepackage{xcolor}
\usepackage{url}
\usepackage{pifont}
\usepackage{enumitem}
\usepackage{makecell}
\usepackage{flushend}
\usepackage{subfigure}

\def\BibTeX{{\rm B\kern-.05em{\sc i\kern-.025em b}\kern-.08em
    T\kern-.1667em\lower.7ex\hbox{E}\kern-.125emX}}
\begin{document}

\title{Logzip: Extracting Hidden Structures via Iterative Clustering for Log Compression}


    
\author{
\IEEEauthorblockN{Jinyang Liu\IEEEauthorrefmark{6}\IEEEauthorrefmark{2}, Jieming Zhu\IEEEauthorrefmark{5}, Shilin He\IEEEauthorrefmark{2}, 
Pinjia He\IEEEauthorrefmark{4}\IEEEauthorrefmark{1}\thanks{\hspace{-2ex}\IEEEauthorrefmark{1}Pinjia He is the  corresponding author.}, Zibin Zheng\IEEEauthorrefmark{6}, Michael R. Lyu\IEEEauthorrefmark{2}\vspace{2ex}}

\IEEEauthorblockA{
    \IEEEauthorrefmark{6}Sun Yat-Sen University, Guangzhou, China~~~~
    \IEEEauthorrefmark{5}Huawei Noah's Ark Lab, Shenzhen, China\\
    \IEEEauthorrefmark{2}The Chinese University of Hong Kong, Hong Kong, China~~~~\IEEEauthorrefmark{4}ETH Zurich, Switzerland\\
    liujy57@mail2.sysu.edu.cn,
    ~~jmzhu@ieee.org,
    ~~\{slhe, lyu\}@cse.cuhk.edu.hk, \\~~pinjia.he@inf.ethz.ch, ~~zhzibin@mail.sysu.edu.cn
}
}



\maketitle

\begin{abstract}
System logs record detailed runtime information of software systems and are used as the main data source for many tasks around software engineering. As modern software systems are evolving into large scale and complex structures, logs have become one type of fast-growing big data in industry. In particular, such logs often need to be stored for a long time in practice (e.g., a year), in order to analyze recurrent problems or track security issues. However, archiving logs consumes a large amount of storage space and computing resources, which in turn incurs high operational cost. Data compression is essential to reduce the cost of log storage. Traditional compression tools (e.g., gzip) work well for general texts, but are not tailed for system logs. In this paper, we propose a novel and effective log compression method, namely \texttt{logzip}. Logzip is capable of extracting hidden structures from raw logs via fast iterative clustering and further generating coherent intermediate representations that allow for more effective compression. We evaluate logzip on five large log datasets of different system types, with a total of 63.6 GB in size. The results show that logzip can save about half of the storage space on average over traditional compression tools. Meanwhile, the design of logzip is highly parallel and only incurs negligible overhead. In addition, we share our industrial experience of applying logzip to Huawei's real products.

\end{abstract}


\section{Introduction}\label{sec:intro}

System logs typically comprise a series of log messages, each recording a specific event or state during the execution of both user applications and components of a large system. These logs have widespread use in many software engineering tasks. They are not only critical for system operators to diagnose runtime failures \cite{ll-43, ZhangMRLY17}, to identify performance bottlenecks \cite{Chow_osdi_2014,ll-33}, and to detect security issues \cite{DeepLog,Oprea_dsn_2015}, but also potentially valuable for service providers to track usage statistics and to predict market trends~\cite{LeeLLLR12,Oliner_cacm_2012}. 

Nowadays, logs have become one type of fast-growing big data in industry~\cite{MiranskyyHCL16}. As systems grow in scale and complexity, logs are being generated at an ever-increasing rate. For example, either in the cloud side (e.g., a data center hosts thousands of machines) or in the client side (e.g., a smartphone vendor with millions of smart devices worldwide), it is common for these systems to generate tens of TBs of logs in a single day~\cite{manylogs}. The massive logs could easily lead to several PBs of data growth a year. In addition, each log is usually replicated into several copies, such as in HDFS, for storage resilience. Some important parts of log data are even synchronized across at least two separate data centers for disaster recovery. This imposes severe pressure on the capacity of storage systems.

What's more, many logs require long-term storage, usually a year or more according to the development lifecycle of software products. Historical logs are amenable to discovering fault patterns and identifying recurrent problems~\cite{LimLZFTLDZ14}. For example, many users often rediscover old problems because they have not installed fix packs~\cite{MiranskyyHCL16}. Meanwhile, auditing logs, which record sensitive operations performed by users and administrators, are often required to be kept for at least two years for possible tracking of system misuse in future. Although storage has become much cheaper than before, archiving logs in such a huge volume is still quite costly. It not only takes up a great amount of storage space and electrical power, but also consumes network bandwidth for transmission and replication. 

To reduce the heavy storage cost of log data, our engineering team seeks two directions of data reduction: 1) reducing logs from the source, and 2) log compression. Logs can be largely reduced by requesting developers to print less logging statements and setting appropriate verbosity levels (e.g., \textit{INFO} and \textit{ERROR}). Yet, logging too little might miss some key information and result in unintended consequences~\cite{zhu2015learning,li2017log}. How to set up an optimal logging standard is still an open problem~\cite{FuZHLDLZX14,ZhaoRLSYZ17}. Instead, we focus on log compression in this paper. It is a common practice to apply compression before storing the data on disks. Mainstream compression schemes (e.g., gzip and bzip) can usually reduce the size of logs by a factor of 10~\cite{MLC-11}. These general-purpose compression algorithms allow for encoding arbitrary binary sequences, but can only exploit redundant information within a short sliding window (e.g., 32KB in gzip's Deflate algorithm). As such, they cannot take advantage of the inherent structure of log messages that might enable more effective compression.

To address this problem, in this paper, we present logzip, a novel log compression method. In contrast to traditional compression methods, logzip can compress large log files with a much higher compression ratio by harnessing the inherent structures of system logs. Log messages are printed by specific logging statements, thus each has a fixed message template. The core idea of logzip is to automatically extract such message templates from raw logs and then structurize them into coherent intermediate representations that are better suitable for general-purpose compression algorithms. To achieve this, we propose the iterative clustering algorithm for structure extraction, following an iterative process of sampling, clustering, and matching. Logzip further generates three-level intermediate representations with field extraction, template extraction, and parameter mapping. These transformed representations are further fed to a traditional compression method for final compression. The whole logzip process is designed to be efficient and highly parallel. As a side effect, the structured intermediate representations of logzip can be directly utilized in many downstream tasks, such as log searching and anomaly detection, without further processing. 

We evaluate logzip on five real-world log datasets (i.e., HDFS, Spark, Andriod, Windows, and Thunderbird) from the loghub repository~\cite{ICSE-benchmark}. They are chosen to span multiple types of systems, including distributed systems, mobile systems, operating systems, and supercomputing systems, and also have different sizes ranging from 1.58 GB to 29.6 GB. The experimental results confirm the effectiveness of logzip, which achieves high compression ratios: 16.2$\sim$813.2. Compared to traditional compression schemes (i.e., gzip, bzip2, lzma), logzip achieves additional 1.3X$\sim$15.1X compression ratios. This leads to a reduction of 47.9\% storage cost on average. Additionally, logzip is highly efficient since the proposed iterative clustering algorithm can be embarrassingly parallelized. We have successfully applied logzip to a real product of Huawei and also share some of our experiences. We emphasize that logzip is generally applicable to all system-generated textual logs, and we leave its use for binary logs for future research. 
 
In summary, our paper makes the following contributions:
\begin{itemize}
 \setlength{\itemsep}{0.5ex}
  \item We propose an effective compression method, logzip, which leverages the hidden structures of logs extracted by iterative clustering.
  \item Extensive experiments are conducted on a range of log datasets to validate the effectiveness and the general applicability of logzip.
  \item We not only share our success story of deploying logzip in industry, but also open the source code of logzip\footnote{https://github.com/logpai/logzip} to allow for future research and practice.
\end{itemize}

The remainder of this paper is organized as follows. Section~\ref{sec:structure} introduces the structure of system logs. We present our iterative structure extraction approach in Section~\ref{sec:ISE} and then describe its use in log compression in Section~\ref{sec:approach}. The experimental results are reported in Section~\ref{sec:experiment}. The industrial case study is described in Section~\ref{sec:case_study}. We review the related work in Section~\ref{sec:relatedwork} and finally conclude the paper in Section~\ref{sec:conclusion}.

\begin{figure}[tbp]
    \centering
    \includegraphics[scale = 0.85]{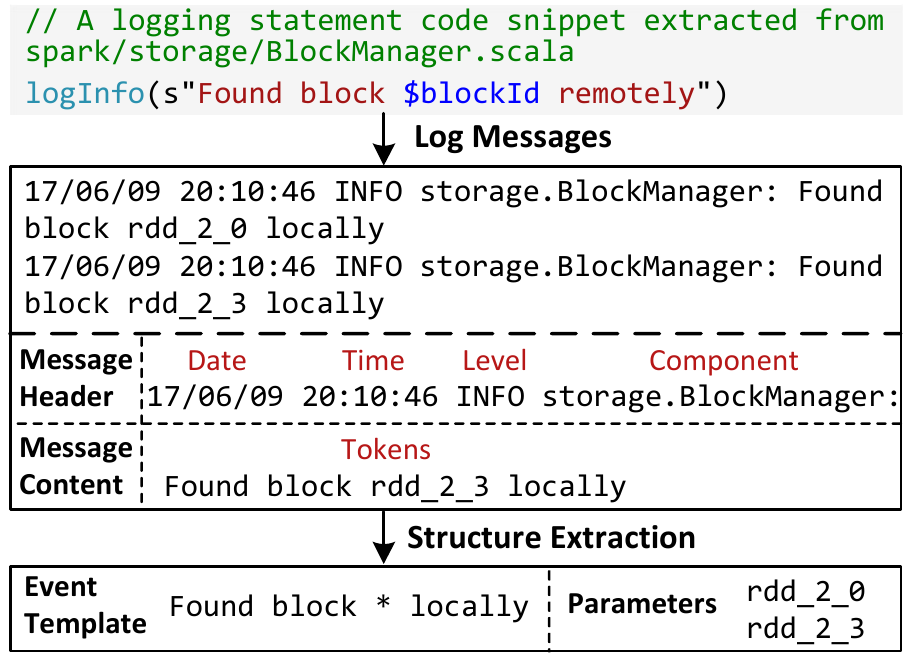}
    \caption{An Example of Extracted Log Structure}
    \label{fig:log_structure}
    \end{figure}


\section{Log Structure}\label{sec:structure}
In this section, we introduce the structures of execution logs, which will be utilized to facilitate log compression.
Fig.~\ref{fig:log_structure} shows the outputs of the logging statement logInfo(s``Found block \$blockId locally'') in the source code of Spark. ``logInfo'' is a logging framework in Scala, and the free text within the brackets are written by developers. This logging framework automatically records information like logging date, time, verbosity level, component, etc. When the logging statement is executed, it outputs a log line like ``17/06/09 20:10:46 INFO storage.BlockManager: Found block rdd\_2\_0 locally.''. The cluster-computing framework Spark uses logs like this to monitor its execution. In practice, a large-scale software system such as Spark records a great deal of information. To reduce the storage cost, we focus on optimizing compression of logs by exploring the structure of logs.


Specifically, hidden structures can be observed in the example in Fig.~\ref{fig:log_structure}. The log message contains two parts, the \textit{message header} automatically generated by the logging framework, and the \textit{message content} recorded by developers.

There are several fields in the message header, such as ``Date", ``Time", ``Level", ``Component". The format is generally fixed for a system since developers barely change the logging framework they use.
Therefore, it is possible to easily extract these fields from each log of a system by using manually defined regular expressions. If more than one logging frameworks are involved, users could define different regular expressions according to different log formats, which only takes minutes for a developer.

Unlike the message header, the message content is unstructured because developers are allowed to write free-form texts to describe system operations.
However, it is possible to find hidden structures in the message content. For example, in Fig~\ref{fig:log_structure}, ``\$blockId'' in the logging statement is a variable that may change in every execution (i.e., \textit{variable part}), whereas other parts remain unchanged (i.e., \textit{constant part}).
We propose to automatically extract the constant part from raw logs as hidden structure via iterative structure extraction (ISE) . 
In the process, the constant part and variable part of a given raw log message can be distinguished. In this paper we denote the constant part as event template (or template in short), and the variable part as parameters.

\section{Iterative Structure Extraction}\label{sec:ISE}
Our proposed approach logzip mainly leverages the hidden structures of logs to facilitate log compression. In this section, we introduce the iterative clustering algorithm for hidden structure extraction.  

\subsection{Overview}

There are three major ways to extract templates\footnote{We use hidden structure and template interchangeably in this paper.} from logs: (1) manual construction from logs; (2) extraction from logging statements in source code; and (3) extraction from raw logs.
In practice, software logs have complex hidden structures and are large-scale. Thus, manual construction of templates is labor-intensive and error-prone. Additionally, the source code of specific components of the system is often inaccessible (e.g., third-party libraries). Therefore, template extraction from software logs is the most widely-applicable approach and thus logzip proposes an iterative clustering algorithm to extract templates from logs automatically.
According to the benchmark by Zhu et al~\cite{ICSE-benchmark}, existing template extraction approaches could perform accurately on software logs. 
However, these methods require all the historical logs as input, leading to severe inefficiency and hindering them from adoption in practice.
Inspired by the cascading clustering by He et al. \cite{shilin-fse}, we propose iterative structure extraction (ISE), which effectively extracts templates from only a fraction of the historical logs.

\begin{figure}[tbp]
\centering
\includegraphics[scale = 0.7]{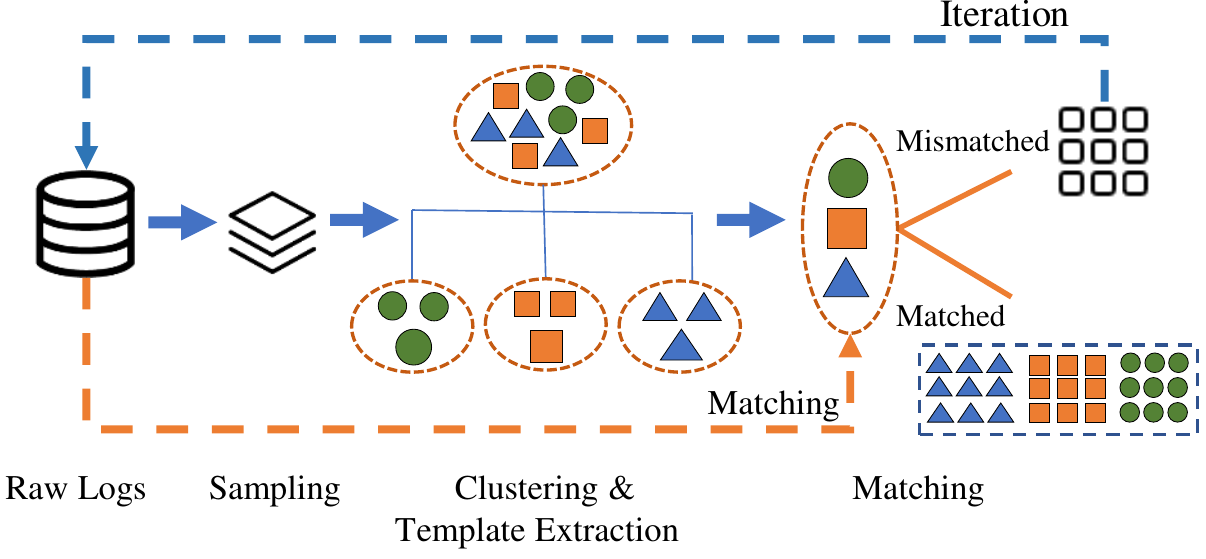}
\caption{Overview of Iterative Structure Extraction}
\label{fig:workflow}
\vspace{-1em}
\end{figure}

Figure \ref{fig:workflow} illustrates the overview of ISE. ISE is an iterative algorithm containing 3 steps in each iteration: sampling, clustering, and matching. The input of ISE is a log file consisting of raw log messages, and the output is extracted templates and structured logs. Specifically, in an iteration, we first sample a portion of the input logs. A hierarchical division method is then applied to the sample logs to generate multiple clusters, from which templates can be extracted automatically. In the matching step, we try to match all the unsampled raw logs with these templates, collect unmatched logs, and feed them into the next iteration as input. By iterating these steps, all log messages could be matched accurately and efficiently with a proper template assignment. The reason behind this is that the sampled logs can often cover the templates hidden in most of the input logs in each iteration. In particular, a fraction of the logging statements could be executed much more frequently than the others. Therefore, templates generated from a small portion of logs can generally match most raw logs at the first several iterations. In the following, we introduce each step of ISE in detail.

\subsection{Sampling}
We first randomly sample a portion of logs from the given raw log file with a ratio $p$. Thus, each log line has an equal probability $p$ (e.g., 0.01) to be selected. If the input contains $L$ log lines, the sampling step results in $S = p\times L$ sampled log lines. This step is inspired by the insight that dominant templates in the original input logs are still dominant in the sampled logs. 


\subsection{Clustering}
These sampled log lines are then grouped into clusters. ISE extracts a template from each cluster by hierarchical divisive clustering in a top-down manner, where we start with a single cluster that consists of all sampled log lines. 
We observed that execution logs have multiple features (i.e., verbosity level, component name, and most frequent tokens) that can be utilized to distinguish the clusters they belong to. Thus, we hierarchically divide logs into coarse-grained clusters by using one feature at each division. 
After that, an efficient clustering algorithm is applied to each of these clusters to further divide logs into fine-grained clusters.
To facilitate efficient log compression, the fine-grained clustering algorithm is designed to be highly parallel. 
We detail the coarse-grained clustering (i.e., divisions by level, component name, most frequent tokens) and the fine-grained clustering algorithms as follows:

\subsubsection{Divide by level}
Intuitively, logs in the same cluster should share the same level, e.g., INFO logs are generally quite different from DEBUG logs because they are recorded for different purposes. Therefore, we first divide logs into clusters according to their levels. 

\subsubsection{Divide by component name}
Similar to the reason for using the level feature, logs generated by different components in a system are barely in the same cluster. So we further separate logs with the same level by their components.

\subsubsection{Divide by frequent tokens}
Intuitively, the constant parts of a log generally have higher occurrences than its parameter parts, because the parameter parts may vary in executions of a logging statement while the constant parts do not. Therefore, it is reasonable to group logs that share the same frequent tokens into the same cluster.
To achieve this, we first tokenize each log message to a list of tokens by using system defined (or as user input) delimiters (e.g., comma and space).
Then, we count the frequency of each token in the sampled logs. 
After that, we find the top-1 frequent token for each log line, according to which we further divide the clusters obtained from the last division using component names. Thus, in this step, logs grouped to the same cluster share the same top-1 frequent tokens.
Moreover, top-2, top-3,..top-$N$ frequent tokens can be applied in the same way to further divide the clusters, where $N$ is a tunable parameter that is normally set to 3.

\begin{figure}[tbp]
\centering
\includegraphics[scale = 1.1]{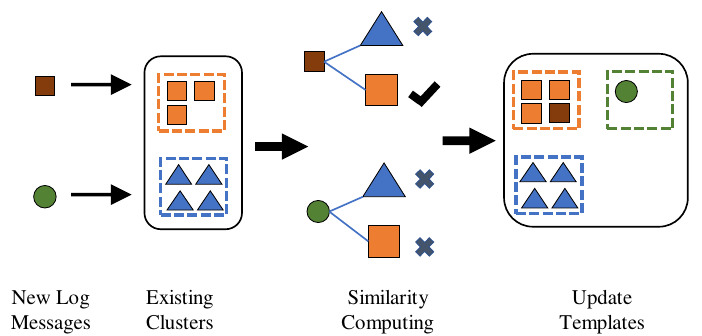}
\caption{Workflow of Sequential Clustering}
\label{fig:clustering}
\vspace{-1em}
\end{figure}

\subsubsection{Divide by fine-grained clustering}
The hierarchical division by log features results in coarse-grained clusters. Logs in the same cluster share the same features as described above. However, these features are not sufficient to determine fine-grained clusters from which we could extract accurate templates. Therefore, we further conduct fine-grained clustering on each of the clusters. 
Inspired by Spell~\cite{spell}, we use longest common subsequence (LCS) to compute the similarity between log messages. Importantly, we also improve the original LCS for speedup. We define the improved similarity as 
$$\phi(a,b)=|a\cap b|$$
where $a$ and $b$ are tokenized log messages, $|\cdot|$ denotes the number of tokens in a sequence. In other words, $\phi(a,b)$ is the number of common tokens of a and b.

We perform the fine-grained clustering in a streaming manner. Fig.~\ref{fig:clustering} describes the workflow of our method. Given a log message $m$, we first tokenize it, then assign it to an existing cluster. To be more specific, we compute the similarity between the input log message with the representative template of each existing cluster, while we keep the largest similarity and the corresponding cluster. If the kept similarity is greater than a threshold of $\theta$, we assign the input log message to the cluster. Note that $\theta=|m|/2$ by default, where $|m|$ denotes the number of tokens contained in the input log. After the assignment, we update the template of the cluster as $LCS(m,t)$, where $t$ is the old template representing the cluster. Note when computing LCS, we mark ``*'' at the places where the two sequences disagree.
For example, the LCS of the two logs ``Delete block: blk-231, blk-12'' and ``Delete block: blk-76'' is ``Delete block:~*''. If the largest similarity could not reach $\theta$,  we create a cluster for $m$, with $m$ itself as the representative template.

The time-consuming step is the computation of similarity between the given log and each template of existing clusters. We propose to use the number of common tokens instead of LCS to measure similarity, which is much more efficient yet effective for two reasons: (1) logs with same tokens but different orderings rarely occur in logs. (2) we have utilized obvious log features to divide logs into coarse-grained clusters. In each of the clusters, logs are expected to share only a few templates, i.e., there are few opportunities for conflicts to occur.

As described above, the sampled logs are divided into clusters hierarchically, each of which has a template to represent logs within the cluster. We emphasize that the clustering algorithm is highly parallel. In particular, the clusters after each division are independent so they could be dispatched to different nodes for parallel computation on the subsequent steps.

\begin{figure*}[!t]
\centering
\includegraphics[scale = 1.2]{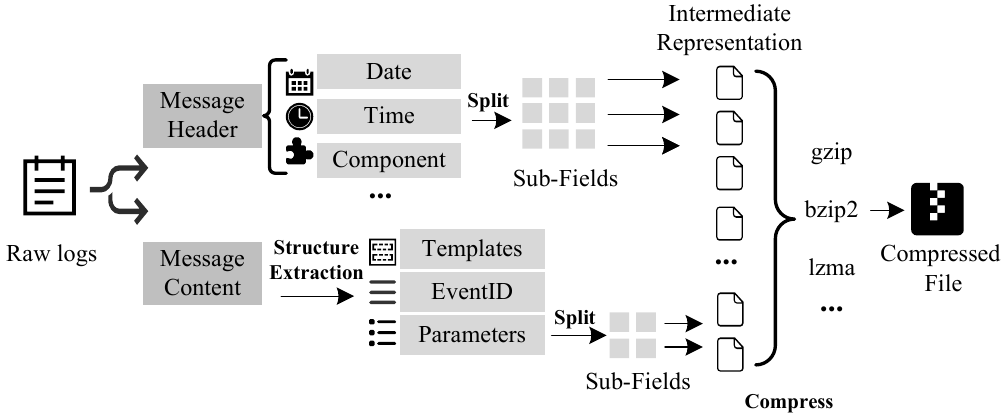}
\caption{Overall Framework of Logzip Compression}
\label{fig:framework}
\end{figure*}
\subsection{Matching}
After collecting all templates from clusters, we use the templates to match each unsampled log message as described in Fig.~\ref{fig:workflow}. By matching, each log message is assigned a template thus the hidden structure is extracted. 
We use the hidden structure to facilitate log compression.

A traditional matching strategy is to transform templates into regular expressions by replacing ``*'' with ``.*?'', then apply regular expressions matching between every combination of log messages and templates, which may explode because of the large number of templates.
To mitigate this efficiency issue, we propose to build all candidate templates as a prefix tree~\cite{wiki-trie}, and perform matching by searching through it.

We build all templates as a prefix tree before searching. The prefix tree starts with a START node. When a template arrives, we tokenize it as is done to a log message. The first token of the log is inserted as a child node of the START node. Then we pass through the token sequence while the previous token is the parent node of the current one. At the end of the last node, we add an END node that contains the whole template for convenience. Intuitively, each template sequence is mapped as a path in the prefix tree. We put all templates into the same tree, and since different templates can have several prefix tokens in common, their paths may overlap.  

Because of sharing the prefix tokens, a log message is compared with all templates at the same time while searching in the compact tree.
Specifically, given a log message, we tokenize it and read from the first token to the last while comparing with nodes in the tree. For the first token, we search if it exists in the second layer of the tree (the first layer is a START node). If the first token matches a node, we continue to check the second token and the children of the node. We stop when all tokens are read. If an END node is reached, we return the template, or we return NONE to denote mismatching. Note that ``*'' in a template denotes parameters with variable length, thus we allow ``*'' in the tree to hold more than one tokens if no child node of ``*'' matches the next log token. For example ``Delete block:``*'' can successfully match ``Delete block: blk-231,  blk-12''. In addition, parameters of a log could be extracted while matching by keeping tokens that match ``*''. In the above example, ``blk-231,  blk-12'' is the parameter. As a result of the matching step, the template and parameters of a log message are extracted if it matches successfully.

The intuition of the tree matching scheme is to compress all templates into a prefix tree by overlapping paths. Therefore, the comparison between log message and all templates becomes one-pass searching. Moreover, checking whether a token matches a node can be done in $O(1)$ by hashing, which makes the tree based strategy a lot more efficient in comparison with regular expression matching. More importantly, the matching step is highly parallel because the search for different log lines on the tree is independent.

\subsection{Iteration}
At the end of each iteration, ISE obtains several templates that could cover all sampled logs. These templates are sufficient to match the majority of all the input log messages in this iteration while some log messages may remain unmatched. Therefore, we repeat the above procedures (i.e., sampling, clustering, and matching) for the unmatched log messages as shown in Fig.~\ref{fig:workflow}. To this end, new templates are extracted from these log messages, and new unmatched data is generated in each iteration. We keep iterating until the percentage of matched log messages reaches a user-defined threshold (empirically, 90\%).

In practice, logging statements of a system evolve slowly. Therefore, ISE could be considered as a one-off procedure for a specific system. To be more specific, we can perform ISE on a portion of logs of the system, and collect templates for future use. After having templates, we could extract structures of new logs from the system through matching instead of running the ISE.

\section{Log Compression}\label{sec:approach}
In this section, we present our log compression method, logzip. We first summarize the workflow of logzip. Then the detail of the compression approach is introduced.

\subsection{Overview of Logzip}
The main idea behind logzip is to reduce the redundant information contained in the original log file.
Fig.~\ref{fig:framework} depicts the overall framework of logzip. For each raw log message in the log data, it is firstly structurized into message header and message content via manually defined regular expressions. 
Then, the message header is split into multiple objects according to their fields and further sub-fields. 
Regarding message content, hidden structures are extracted by applying ISE. After that, we represent each log message as a template, an event ID as well as the corresponding parameter. In addition, each item in the parameter list is split into sub-fields. Then, those logs that share the same event ID are stored into the same object in a compact manner.
At last, all generated objects are compressed to a compact file by existing compression tools. Details are described as follows.

\begin{figure*}[!ht]
\centering
\includegraphics[scale = 0.94]{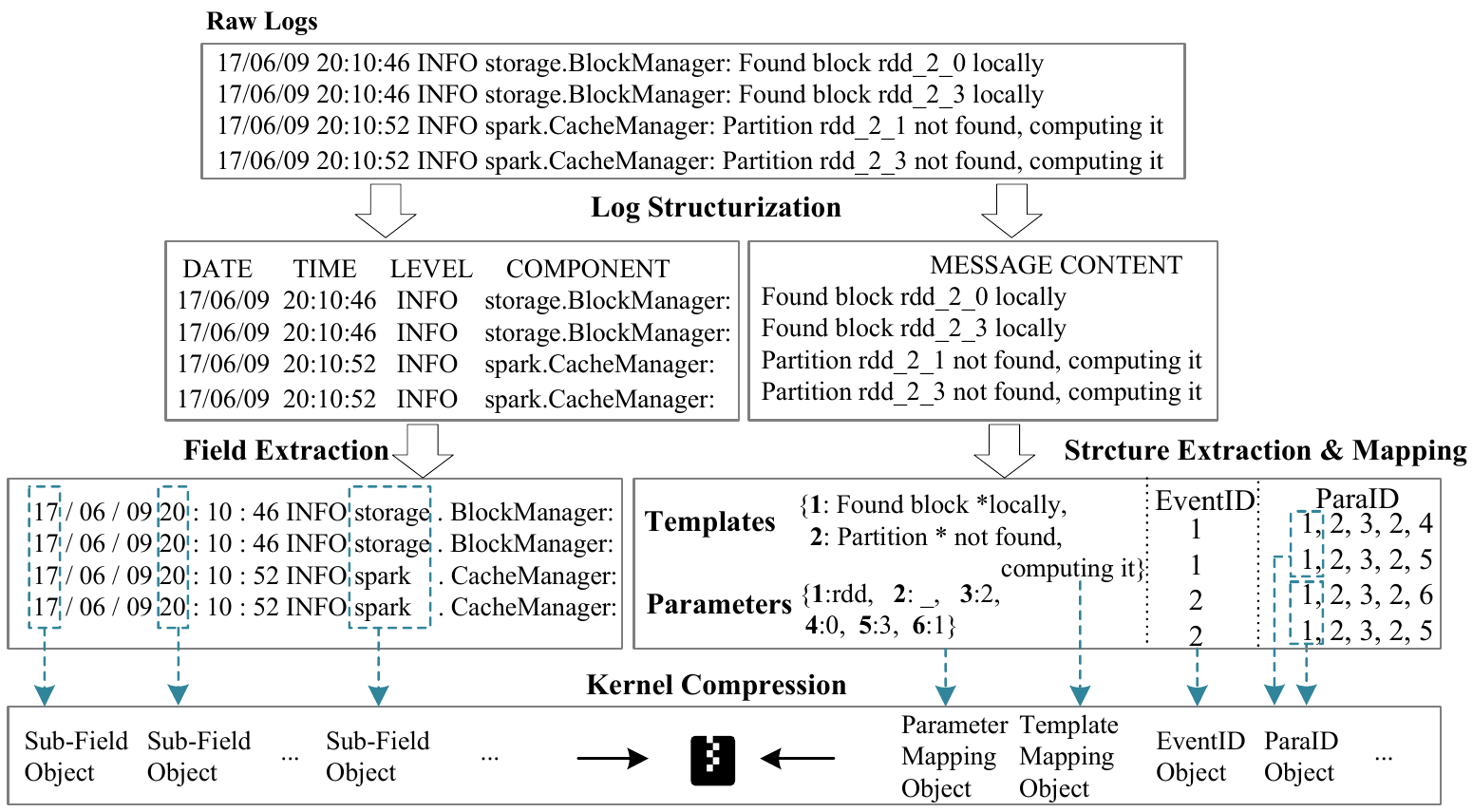}
\caption{An Example of Logzip Workflow in Three Levels}
\label{fig:example}
\end{figure*}

\subsection{Approach}
Logzip can perform compression in 3 levels, and achieve different effectiveness and efficiency. Fig.~\ref{fig:example} is an example of logzip workflow.

\textbf{Level 1: Field Extraction}. We first extract fields of given raw logs by applying a user-defined regular expression. For the example in Fig.~\ref{fig:example}, ``DATA'', ``TIME'' and ``LEVEL'' could be extracted by identifying the white space delimiter. ``COMPONENT'' and ``MESSAGE CONTENT'' are separated by a colon. We emphasize that it is easy to manually construct the regular expression, which generally remains unchanged for a specific system since the message header is automatically recorded by the logging framework (e.g., log4j in Java, logging in Python). Then, each field is further split into sub-field according to special characters (e.g., non-alphabetic characters) to increase the coherence in a sub-field. These sub-fields are then stored into separate objects. For level 1, we do not process the message content and store it into an object. 

\textbf{Level 2: Template Extraction}.  
The message content is further processed by extracting the hidden structures, i.e., message templates.
We directly apply ISE as described in Section~\ref{sec:ISE}. After that, the message content could be represented as its template and parameters, and we assign auto-incremental EventID initialized to 0 for unique templates, which forms a template mapping dictionary (EventID is the key and the corresponding template is the value). 
In fact, a template may be shared by many log messages. For example, The HDFS logs that we studied contain around 11.2 million log messages, but they share only 39 templates.
Therefore, we use the corresponding EventId to denote the template of each log message. 
In doing this, log messages are transformed into a compact form containing short EventId and parameters. At last, the template mapping dictionary is stored into an object alone, while the EventIds are stored in an EventId object. 
For the parameters extracted from the message content, we split each item of parameters in a similar manner as in level 1. Clearly, each parameter is split with non-alphabetic characters as delimiters. Then each generated sub-field within a group constitutes one object separately. Here a group represents all logs that share the same template. The intuition behind is that parameters within a group may be duplicated or similar, and putting similar items into a file could make the best of existing tools, e.g., gzip.

\textbf{Level 3: Parameter Mapping}. We further optimize the representation of parameters in level 3. Based on our observation, some inseparable and very long parameters (i.e., no delimiter inside) waste too much space. For example, the block ID (e.g., ``blk\_-5974833545991408899") is space-consuming and may have high occurrence.
To sidestep the problem, as shown in Fig.~\ref{fig:example}, we encode unique sub-field values to sequential 64-base numbers (ParaID), which forms a parameter mapping dictionary (ParaID is the key and the corresponding parameter is the value). For the sake of saving more space, parameters from all groups share the same parameter mapping dictionary.
To conclude, in level 3, parameters are encoded into ParaIDs. At last, one parameter mapping dictionary object and ParaID objects for each group are generated separately.

\textbf{Compression}. 
After three levels of splitting, encoding and mapping, several objects are generated. The last step is to pack all these objects to be a compressed file without losing any information. Since our main interest lies in the aforementioned three levels of processing, we directly utilize those off-the-shelf compression algorithms and tools in this step, e.g., gzip, bzip2, and lzma. In this way, our logzip is compatible with existing compression tools and algorithms. It is worth noting that most log analysis algorithms (e.g., anomaly detection~\cite{LKE,ll-42}) take templates as input without parameters. Therefore, logzip could perform lossy compression in this case by discarding all parameter objects before compression with existing tools, which could be more effective.

\textbf{Decompression}. 
As the reverse process of log compression, decompression should be able to recover the original dataset without losing any information. At first, multiple objects are generated after unzipping the compressed file. Then, we recover the message header by simply merging all the sub-fields values extracted in level 1 in order. As for recovering the message content, we first get the templates by indexing the event encoding dictionary with the EventID. Similar, the parameter list is retrieved by indexing the parameter encoding dictionary using the ParaID. By replacing the ``*" in the template using parameters in order, the message content can be completely recovered.

\section{Evaluation}\label{sec:experiment}
In this section, we conduct comprehensive experiments by applying logzip to a variety of log datasets and report the results. We aim to answer the following research questions.
\begin{itemize} 
\setlength{\itemindent}{1.8ex}
\setlength{\itemsep}{0.5ex}  
\item[\textit{RQ1}:] What is the effectiveness of logzip?
\item[\textit{RQ2}:] How effective is logzip in different levels?
\item[\textit{RQ3}:] What is the efficiency of logzip?
\end{itemize}

\begin{table}[!ht]
    \centering
    \caption{Summary of Log Datasets } \label{tab:datasets}
    \includegraphics[width=0.48\textwidth]{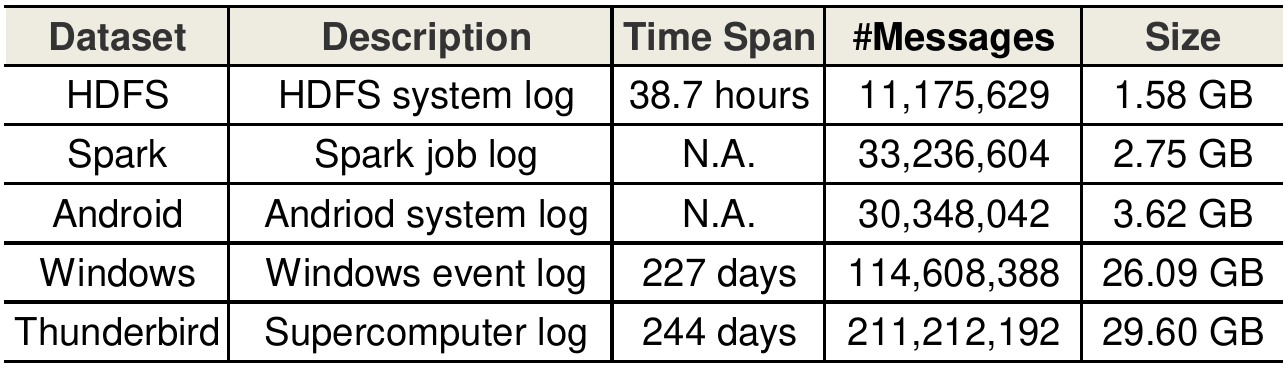}
    \end{table}

\begin{table*}[htbp]
    \centering
    \caption{Compression Results w.r.t. Size (in MB) and Compression Ratio (CR) of Different Compression Methods
    \protect\\
    (Cowic~\cite{cowic} and LogArchive~\cite{logarchive} are baseline algorithms for log compression.)} \label{tab:RQ1}
    \includegraphics[scale=0.75]{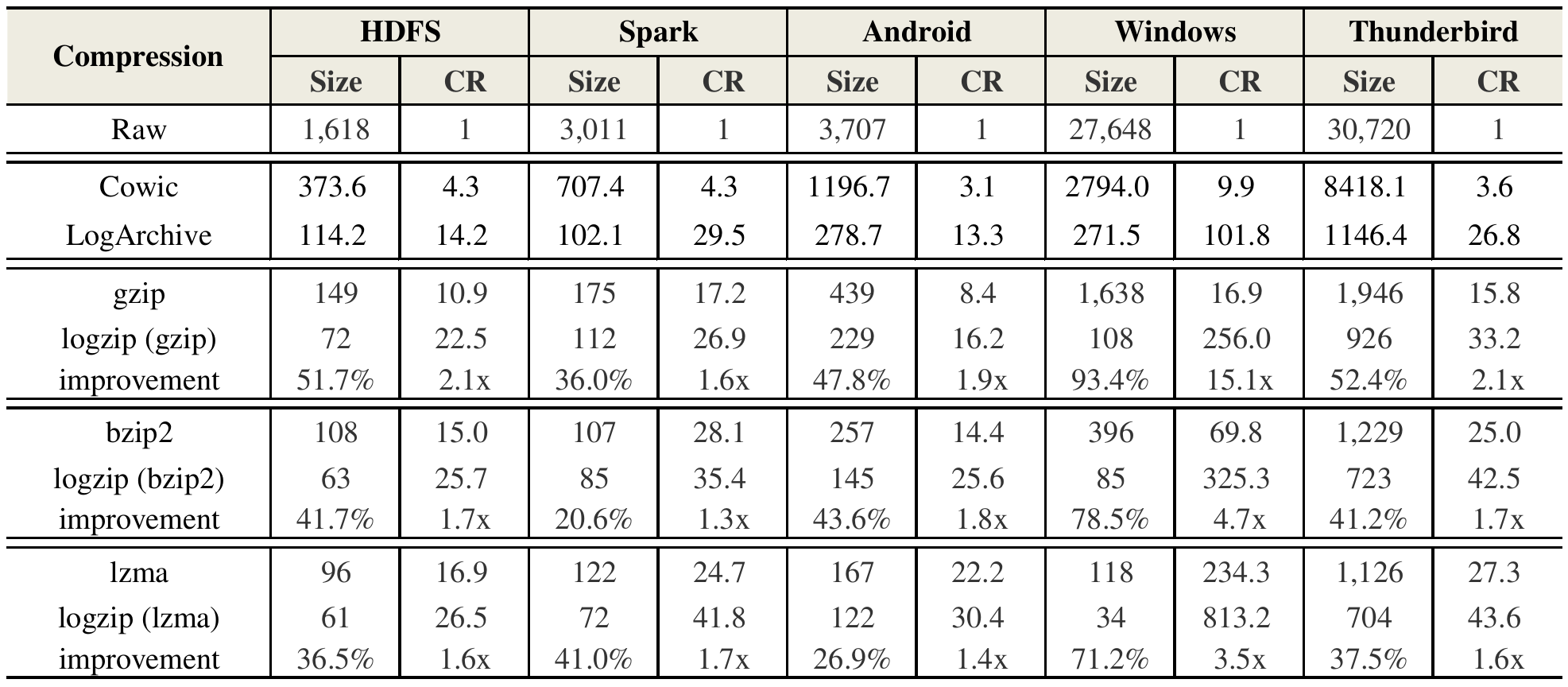}
\end{table*}

\subsection{Experimental Setup}
\textbf{Log Datasets:}
We use five representative log datasets to evaluate logzip, as presented in Table~\ref{tab:datasets}. These log datasets are generated by various systems spanning Distributed Systems (HDFS, Spark), Operating System (Windows), Mobile System (Android) and Supercomputer (Thunderbird). Some datasets (HDFS \cite{Xu_sosp_2009}, Thunderbird \cite{BGLdata}) are released by previous log research, while the other (Spark, Windows, Android) are collected from real systems in our lab environment. Moreover, the total size of all datasets is around 63.6 GB, which contains a total amount of more than 400 million log messages. All the datasets that we use are available on our Github.

\textbf{Evaluation Metrics:} To measure the effectiveness of logzip, we use \textit{Compression Ratio (CR)}, which is widely utilized in the evaluation of compression methods.  The definition is given below:
$$ CR = \frac{Original\ File\ Size}{Compressed\ File \ Size}$$
Note that the original size of a given log dataset is always fixed while the compressed file size may vary. When reducing the compressed file size, a higher compression can be achieved, which indicates more effective compression.

\textbf{Compression Kernels:}
As introduced in Section \ref{sec:approach}, logzip utilizes existing compression utilities as the compression kernel in the last step. In the experiments, three prevalent and effective compression algorithms (i.e., gzip, bzip2, lzma) are selected. Note that these compression algorithms can also be employed to compress log files solely, which will serve as baselines in our experiments.

\textbf{Experimental Environment:}
We run all experiments on a Linux server with Intel Xeon E7-4830 @ 2.20GHZ CPU and 1TB RAM, running Red Hat 4.8.5 with Linux kernel 3.10.0.

\subsection{RQ1: Effectiveness of Logzip}
To study the effectiveness of logzip, we use logzip it to compress all five collected log datasets. As introduced before, we use these existing popular compression tools (i.e., gzip, bzip2, lzma) as well as two log compression algorithms (i.e., Cowic~\cite{cowic}, LogArchive~\cite{logarchive}) as baselines for a fair comparison. Since logzip can be equipped with different compression kernels, it also has three variants, i.e., logzip (gzip), logzip (bzip2), logzip (lzma). We report both the compressed file size and the compression ratio (CR) in Table~ \ref{tab:RQ1}. Note that all results of logzip are obtained in level 3. 

We first make brief comparisons among gzip, bzip2 and lzma. lzma is generally the most effective one on most datasets while gizp performs the worst.
As for the two algorithms specifically designed for log data, Cowic and LogArchive, LogArchive could achieve higher CR than gzip but is generally less effective than bzip2 and lzma. Cowic is even worse than gzip, since Cowic is designed for a quick query on compressed data instead of pursuing high CR.

Logzip variants with different compression kernels result in different compressed size, which is determined by the effectiveness of the kernels.
Compared with the three baseline methods, logzip equipped with the corresponding compression kernel achieves higher CR on all five datasets. In particular, logzip can achieve a CR of 4.56x on average and 15.1x at best over the gzip traditional mature compression algorithm. For example, the compressed file is around 149MB by gzip while 72MB by logzip (gzip), and our method can save around half of the storage, which is crucial in practice. Logzip equipped with other compression kernels achieves similar results. 

\begin{figure*}[htbp]
    \centering
    \includegraphics[width=0.7\textwidth]{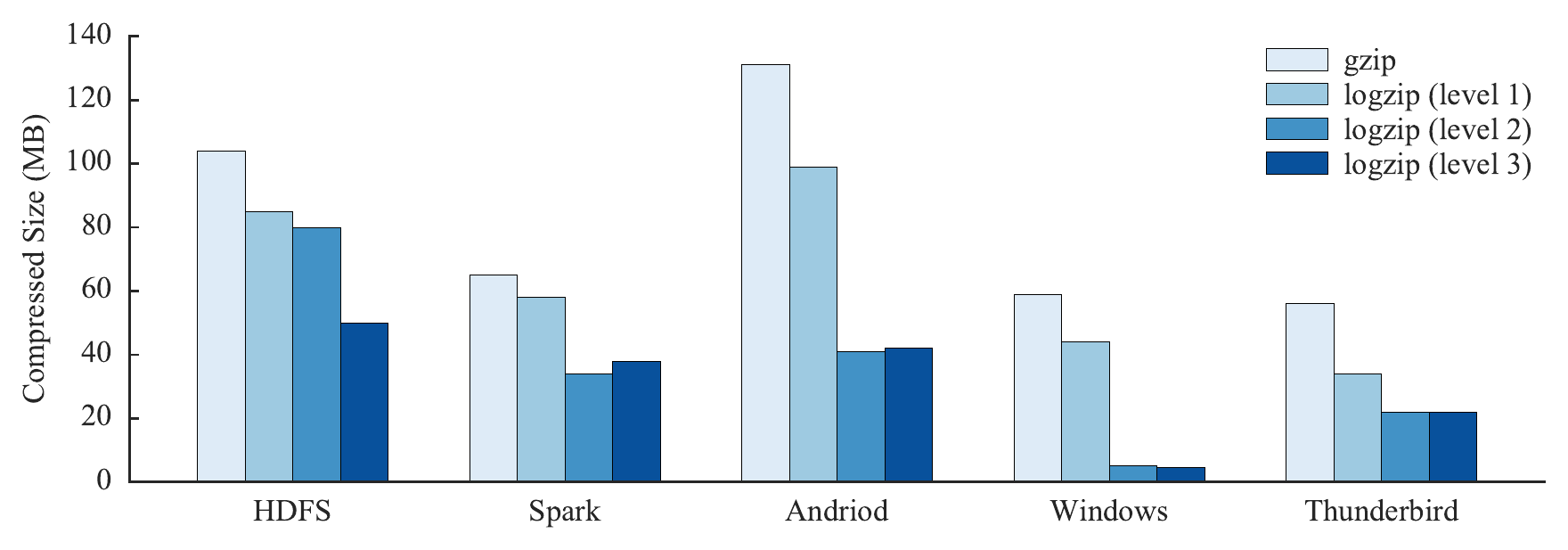} 
     \vspace{-1em}
    \caption{Compressed File Size (MB) in Different Levels} \label{fig:logzip_levels}
    \end{figure*}

\subsection{RQ2: Effectiveness of Logzip in Different Levels}\label{sec:RQ2}

As introduced in Section \ref{sec:approach}, logzip is designed in three levels, splitting the original log message into multiple objects with different fineness. In this section, we evaluate the effectiveness of logzip at each level. 
In practice, logs are generated and collected in a streaming manner, and stored as a file when they grow to a proper size, e.g., 1GB. For the simplicity of comparison, our experiments are conducted on the first 1GB logs of all five datasets.
Besides, since our focus lies in varying different levels and compression kernel is not a major concern, we conduct experiments by taking gzip as the baseline method and logzip (gzip) as our compression approach. Based on this, we vary the level of logzip (gzip) to evaluate the effectiveness of an individual level, and the experimental results should also apply to other compression kernels. 

Fig.~\ref{fig:logzip_levels} shows compressed file sizes on five datasets in different levels. We can find that level 1 field extraction works well on all datasets, compressing the original 1GB log files to files of less than 100MB. It is because generally most fields of a log file have limited unique values, and gzip is able to compress such text data into a file of small size. Moreover, compared to the baseline gzip compression, logzip with only level 1 already achieves much better compression results.

Considering the structure extraction of message content in level 2, it sharply reduces the compressed size on almost every dataset. In particular, after applying logzip (level 2), the compressed log file of Android takes up only $\frac{1}{3}$ of the baseline gzip-compressed file, and similarly, the fraction is even less than $\frac{1}{10}$ on Windows. The results confirm the effectiveness of level 2 in log compression. The reason is also straightforward. In level 2, ISE extracts the invariant templates out of log message content and replace it with an event ID. Hence, only keeping event IDs instead of original template strings saves a large amount of storage space.
Besides, parameters are also split in a similar way as level 1 field extraction, which contributes to reducing the compressed file size. 
However, we also observe that the improvement is not obvious on the HDFS log file. After close analysis on the HDFS logs, we find that the major part of HDFS log message content is parameters instead of the template. Parameters such as ``Block Id'' (e.g., ``blk\_-5974833545991408899'') are too long and cannot be separated in level 2. Therefore, the compressed file size of the HDFS data is not as small as other log files.

In level 3, we map parameters into 64-base numbers. As shown in Fig.~\ref{fig:logzip_levels}, logzip gets at least half the compressed size compared with gzip on all five datasets. 
In particular, comparing to logzip (level 2), the compressed file size is greatly reduced on HDFS dataset, which confirms the importance of encoding the long and duplicate parameters as aforementioned.
Comparable or slightly worse results show on other datasets. This is caused by introducing extra ParaIDs for these logs without many space-consuming parameters. In fact, these extra ParaIDs cost little space, which is tolerable. That is, users could directly apply logzip (level 3) to their dataset to achieve the best performance without considering whether the log contains such parameters.


To conclude, logzip is effective in every level for the log dataset that we study. More importantly, logzip theoretically generalizes well for the text log data of other types for 2 reasons: (1) Only a little prior knowledge is required when formatting raw logs. Once set up, no more manual effort is required unless a system updates greatly. (2) Logs generated by logging statements naturally contain hidden structures. The key step of logzip, ISE, is able to automatically extract the hidden information used for log compression. 
Note that logzip is designed for logs stored in text form, which is the most common case. Those in a binary format are beyond our consideration, and we will explore the case in future work.

\begin{figure*}[!t]
\centering    
\subfigure[HDFS]
{
	\centering      
	\includegraphics[scale=0.425]{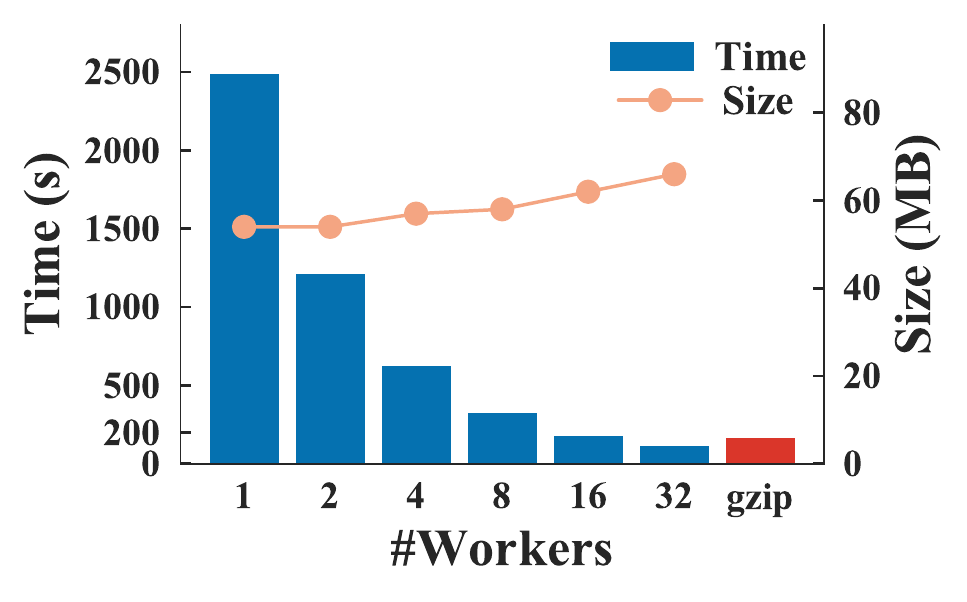}
}
\subfigure[Spark]
{
	\centering      
	\includegraphics[scale=0.425]{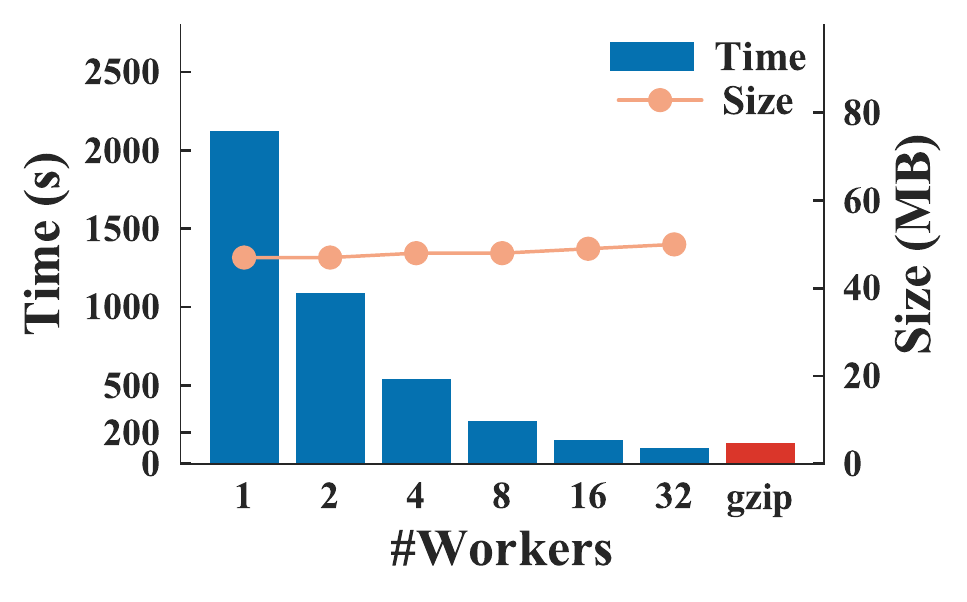} 
}
\subfigure[Thunderbird]
{
	\centering      
	\includegraphics[scale=0.425]{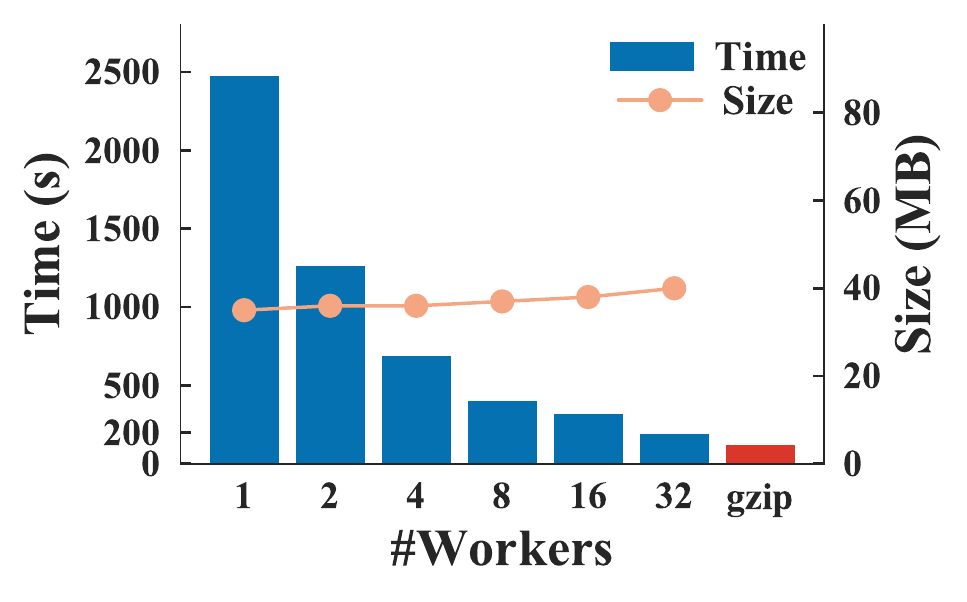} 
}
\subfigure[Windows]
{
	\centering      
	\includegraphics[scale=0.425]{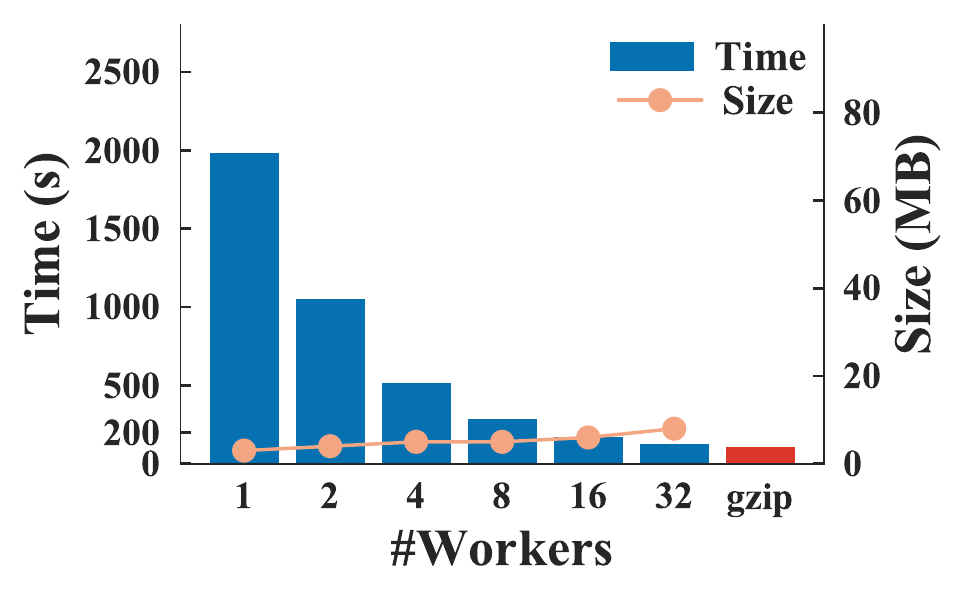} 
}
\caption{Compression Time \& Size vs \#Worker} 
\label{fig:RQ3}  
\end{figure*}
\subsection{RQ3: Efficiency of Logzip}
In this section, we evaluate the efficiency of logzip (gzip) (in short, logzip). Logzip is designed to be highly parallel, thus we would like to know the efficiency achieved by utilizing different numbers of workers. Gzip is known as an efficient compression tool thus used as a baseline. 
We apply both algorithms on the first 1GB log data of HDFS, Spark, Thunderbird and Windows, for the same reason mentioned in section \ref{sec:RQ2}. We exclude Android dataset for saving space. In addition, we vary the worker number of logzip in range $[1,2,4,8,16,32]$.

Fig.~\ref{fig:RQ3} depicts the execution time and compressed size achieved by logzip and gzip, on four datasets. Note that the execution time consists of all steps including ISE (Sec~\ref{sec:ISE}) and compression (Sec~\ref{sec:approach}).
As for logzip, the time cost halved after doubling the number of workers. More
importantly, it is worth noting that the time cost by using 32 workers is comparable with that of gzip, even better in HDFS and Spark. The result shows the parallelizability and efficiency of logzip, which can be explained by the design of logzip: (1) We extract high-quality templates from only a small portion of logs in ISE, e.g., 1\% log messages are sufficient to generate templates that match 90\%$\sim$100\% logs messages. (2) The top-down manner clustering are higly parallel since fine-grained clustering could be  performed simultaneously and independently. (3) We build all templates into a compact prefix tree, the one-pass matching scheme is efficient. (4) A log file could be split into several chunks, and performing logzip on each chunk at the same time is possible to reduce time consumption.

In addition, the compressed size slightly increases with more workers involved. This is the result of chunking of logs. A whole log file is split into chunks before feeding to a worker, as a result, each worker only sees a part of the data, which slightly hinders logzip to utilize the global information for compression.

To conclude, logzip is highly parallel and could achieve comparable or better efficiency than gzip. Note that, after chucking the input log file, the compressed size may increase a little bit, but it is tolerable in practice.

\section{Industrial Case Study}\label{sec:case_study}
At Huawei, logs are continuously collected during the whole product lifecycle. With the rapid growth of scale and complexity of industrial systems, logs become a representative type of big data for software engineering teams at Huawei. For example, System X (anonymized for confidential issues), which is one of the popular products at Huawei, generates about 2 TB of log data daily. Storage of logs at such a scale has become a challenging task. Most of the logs need to be stored for a long time, usually 1$\sim$2 years, considering the product lifecycle of System X is about two years. In particular, historical logs are kept for the following practical tasks: 1) \textit{root cause analysis}: identification of similar failures or faults that happened before; 2) \textit{failure categorization}: categorization of similar failures for the development planning of next software version; and 3) \textit{automated log analysis tools}: acting as experimental data for the research and development of automated log analysis tools.

Log storage currently takes up over PBs of disk space in a cluster, which indicates a huge cost of power consumption. Meanwhile, log data is regularly replicated to one or two data centers (according to different importance levels) at different locations for disaster tolerance. This results in another type of expensive cost, i.e. bandwidth consumption. Reducing the storage cost of log data has become a main objective of the product team because their storage budget is limited but the number of products is growing. With close collaboration with the product team, we have recently transferred logzip into System X. Logzip is deployed on a 64-core Linux server with Ubuntu 16.04 installed to compress raw log files. When logzip is parallelized with 32 processors, it achieves comparable compression time with the traditional gzip method. The product team accepts the performance of logzip, since most of the old logs are rarely accessed and can be archived at one time. Yet, the use of logzip successfully reduces the size of logs, saving about 40\% of space compared to the gzip algorithm that is previously used. It not only reduces the cost of log storage but also cuts the cost on network consumption during replication. This has become a successful use case of logzip. 


\section{Related Work}\label{sec:relatedwork}

\textbf{Log Management for SE.}
Logs are critical runtime information recorded by developers, which are widely analyzed for all sorts of tasks.
1) \textit{Log analysis} is conducted for various targets, such as code testing~\cite{ASE-codecoverage}, problem identification~\cite{ICSE-problem-ident, ICSE-anti-patterns}, user behavior analysis~\cite{Drain-10, Drain-11}, security monitoring\cite{ll-32}, etc.
Most of these tasks use data mining models to extract critical features or patterns from a large volume of software logs. Therefore, we believe our work on log compression could benefit log data storage and save the cost of dumping the large volume of logs.
2) \textit{Log parsing.} is generally utilized as the first step of downstream tasks. In recent years, various log parsers have been proposed. 
SLCT~\cite{SLCT} is the first work on automated log parsing based on token frequency, to the best of our knowledge. Then data mining-based methods (LKE~\cite{LKE}, IPLoM~\cite{IPLoM}, Spell~\cite{spell}, Drain\cite{Drain, Drain-journal}) are proposed. LKE and IPLoM are offline parsers, and SHISO and Drain could parse online in a streaming manner. These parsers are evaluated and compared in the benchmark by Zhu et al.~\cite{ICSE-benchmark}. The parsers could extract hidden structures but they take all logs as input thus are not efficient compared with the proposed ISE. 
%




\textbf{Text Compression.}
File compression algorithms have been developed for years~\cite{MLC-22, MLC-23}, and some of them are utilized in compressing tools (gzip, lzma, bzip2). These general tools are commonly used and achieve satisfactory CR. To further improve CR specific for text files, Lempel-Ziv-Welch (LZW) based methods are widely studied~\cite{SMS, Lempel-Ziv-style, DNA}.
Oswald et al.~\cite{pattern-mining} explore text compression in the perspective of Data Mining. They enhance Huffman Encoding by frequent pattern mining. Since log data as a kind of text data is more structured, we propose logzip to use the information to facilitate compression.





\textbf{Log Compression.}
Due to the inherent structure of log data, it's possible to compress log files with higher CR, thus some log-specific algorithms are proposed. CLC~\cite{MLC-9} and DSLC~\cite{MLC-18} utilize prior knowledge and manual pre-treatment to compress log files. LogArchive~\cite{logarchive} adaptively distributes log messages to different buckets and track most recent log messages in each bucket with a sliding window. Finally, buckets are compressed separately in parallel.
Cowic introduced by Lin et al.~\cite{MLC-11} divides log messages into fields and manually build a model for each field, but Cowic targets query efficiency instead of high CR.
MLC~\cite{MLC} explores data redundancy of log file and divides logs into buckets based on similarities, then apply existing compression tools as we do in logzip.
These algorithms explore hidden structures of logs to compress logs, which could outperform general compression tools. But they are limited by the trade-off between high CR and efficiency. 
Compared with them, we extract hidden structures via ISE, which is efficient and highly parallel. As a result, logzip achieves high CR without loss of efficiency.


\textbf{Log Analytics Powered by AI (LogPAI).} 
LogPAI is a research project originating from CUHK. The ultimate goal of LogPAI is to build an open-source AI platform for automated log analysis. Towards this goal, we have built open benchmarks over a set of research work as well as release open datasets and tools for log analysis research. In particular, loghub hosts a large collection of system log datasets. Logparser provides a toolkit and benchmarks for automated log parsing~\cite{ICSE-benchmark, pinjia-TDSC, He-DSN}. Loglizer implements a number of machine-learning based log analysis techniques for automated anomaly detection~\cite{shilin-fse, he-issre}. LogAdvisor is a framework for determining optimal logging points in source code~\cite{ASE-log-nlp, zhu2015learning,FuZHLDLZX14}. In this work, logzip provides an tool for effective log compression. With both datasets and source code available, we hope that our LogPAI project could benefit both researchers and practitioners in the community.

\section{Conclusion}\label{sec:conclusion}

In this paper, we propose logzip, a log compression approach that largely reduces the operational cost for log storage. Logzip extracts and utilizes the inherent structures of logs via a novel iterative clustering technique. Logzip is designed to be seamlessly integrated with the existing data compression utilities (e.g., gzip). Furthermore, the semi-structure intermediate representations generated by logzip can be directly used for a variety of downstream log mining tasks (e.g., anomaly detection). Extensive experiments on five real-world system log datasets have been conducted to evaluate the effectiveness of logzip. The experimental results show that logzip significantly enhances the compression ratios over three widely-used data compression tools and also outperforms the state-of-the-art log compression approaches. Moreover, logzip is highly parallel and achieves comparable efficiency as gzip on a 32-core machine.  We believe that our work, together with the open-source logzip tool, could benefit engineering teams facing the same problem. 

\section{Acknowledgement}\label{sec:Acknowledgement}

The work described in this paper was supported by the National Key Research and Development Program (2016YFB1000101), the National Natural Science Foundation of China (61722214), the Research Grants Council of the Hong Kong Special Administrative Region, China (No. CUHK 14210717 of the General Research Fund), and Microsoft Research Asia (2018 Microsoft Research Asia Collaborative Research Award).



\bibliographystyle{IEEEtran}
\bibliography{main}

\end{document}